# The Joint Center for Energy Storage Research: A New Paradigm for Battery Research and Development

George Crabtree

*Joint Center for Energy Storage Research, Argonne National Laboratory, 9700 S. Cass Avenue, Argonne, IL 60439, and University of Illinois at Chicago, 845 W. Taylor Street, Chicago IL 60607*

**Abstract.** The Joint Center for Energy Storage Research (JCESR) seeks transformational change in transportation and the electricity grid driven by next generation high performance, low cost electricity storage. To pursue this transformative vision JCESR introduces a new paradigm for battery research: integrating discovery science, battery design, research prototyping and manufacturing collaboration in a single highly interactive organization. This new paradigm will accelerate the pace of discovery and innovation and reduce the time from conceptualization to commercialization. JCESR applies its new paradigm exclusively to beyond-lithium-ion batteries, a vast, rich and largely unexplored frontier. This review presents JCESR's motivation, vision, mission, intended outcomes or legacies and first year accomplishments.

**PACS:** 61, 66, 68, 71, 72, 73, 81, 82, 88

## OVERVIEW

Transportation and the electricity grid account for two-thirds of U.S. energy use [1]. Each of these sectors is poised for transformation driven by high performance, low cost electricity storage. The Joint Center for Energy Storage Research (JCESR) pursues discovery, design, prototyping and commercialization of next generation batteries that will realize these transformational changes. High performance, low cost electricity storage will transform transportation through widespread deployment of electric vehicles; it will transform the electricity grid through high penetration of renewable wind and solar electricity and a new era of grid operation free of the century-old constraint of matching instantaneous electricity generation to instantaneous demand. It is unusual to find transformational change in the two largest energy sectors driven by a single innovation: high performance, low cost energy storage.

These transformative outcomes for transportation and the electricity grid require electricity storage with five times the performance at one-fifth the cost of present generation commercial batteries [2]. The remarkable advances in the present generation of lithium-ion batteries, performance improvements of 8% per year and reductions in cost of 5% per year, cannot reach the factors of five advances that JCESR seeks for transformative change. JCESR looks exclusively beyond lithium-ion for its next-generation electricity storage technologies.

## ENERGY STORAGE CONCEPTS

Lithium ion batteries, conceptualized in the early 1970s and commercialized in the early 1990s [3], store energy by intercalation of singly charged lithium ions in a graphite anode, and release energy by transferring these ions through an organic electrolyte to a lower energy intercalated state in a metal oxide or metal phosphate cathode [4] (Fig. 1). Each positive lithium ion moving from anode to cathode inside the battery is accompanied by a negative electron moving the same direction outside the battery to maintain charge neutrality. This simple conceptual system—lithium intercalation at anode and cathode—can realize significant incremental improvements by substituting higher capacity, higher voltage or faster charging intercalation materials for the electrodes now in use, such as silicon for graphite at the anode and mixed transition metal oxides for pure metal oxide at the cathode [5-13]. The battery community vigorously pursues these research opportunities because the incremental improvements they can achieve in performance, cost and safety are in high demand by device manufacturers and consumers.

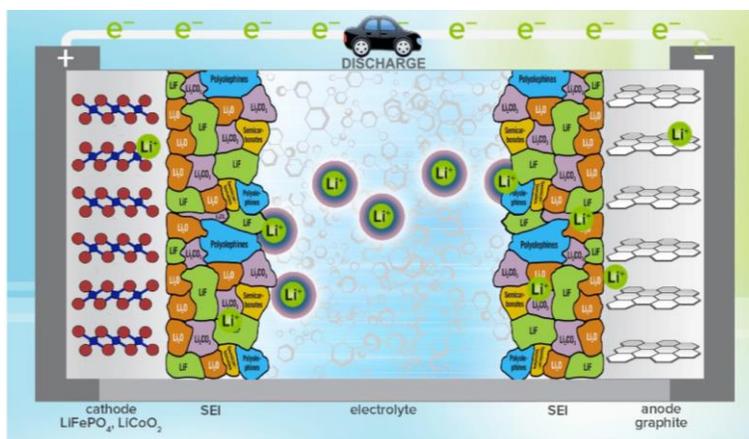

**FIGURE 1.** The present generation of lithium-ion batteries operates with a simple concept: singly charged lithium ions transfer from a graphite anode through an organic electrolyte to a metal oxide or metal phosphate cathode on discharge, and in the reverse direction on charge. For each positively charged lithium ion moving inside the battery, a negatively charged electron moves through the external circuit to maintain charge neutrality.

In contrast, the beyond-lithium-ion battery space is much larger, richer, and far less explored than the lithium-ion space. Unlike the single concept of intercalation at the anode and cathode in commercial lithium ion batteries, beyond-lithium-ion batteries embrace a wealth of new concepts, including multiply charged working ions such as doubly charged magnesium in place of singly charged lithium [14,15,16], high energy covalent chemical reactions at the anode and cathode in place of intercalation [17-24], and fluid electrodes with large storage capacity and low cost in place of crystalline electrodes [25-31].

These three energy storage concepts, multivalent intercalation, chemical transformation, and non-aqueous redox flow, are illustrated in Fig. 2. They are fundamentally different from the intercalation concept of commercial lithium ion batteries, have the potential to reach factors of five higher performance and lower cost, and provide significantly greater design and operational flexibility and opportunity than commercial lithium ion batteries. These three energy storage concepts are the primary research directions that JCESR pursues.

## BEYOND LITHIUM-ION BATTERIES

The scope and complexity of the beyond-lithium-ion space dramatically exceeds that of the lithium-ion space as illustrated in Fig. 3. Mixing and matching JCESR's three energy storage concepts and including the options of liquid or solid electrolytes, illustrated on the vertical axis of Fig. 3, yields a rich spectrum of at least 18 different conceptual designs for beyond lithium-ion battery systems. Add to this the range of 20-30 candidate materials (illustrated on the horizontal axis of Fig. 3) that could implement these battery design concepts and there are 50-100 beyond lithium-ion battery combinations whose feasibility and performance remain to be explored. In contrast, the lithium-ion space has a single storage concept, lithium intercalation at anode and cathode, and fewer than ten materials in play to implement this concept.

Remarkably, the much richer beyond lithium-ion space has far fewer researchers exploring its opportunities than the lithium-ion space. One reason for this disparity is the established market that lithium-ion batteries represent. A 5% improvement in performance or reduction in cost can profoundly influence this large market and produce high returns for an incremental advance. A second reason for the disparity in research activity is the high risk of creating an entirely new technology. Inventing a beyond-lithium-ion battery meeting JCESR's transformative performance goals requires discovering three new materials (one each for anode, electrolyte and cathode), each of which performs five times better than the corresponding lithium-ion material, and which are all electrochemically compatible with each other. The last requirement, mutual compatibility, requires that the battery be created from whole cloth. It cannot be divided into three separate materials tasks that are pursued independently. This is a daunting challenge, beyond the resources of most research groups and small-to-medium sized companies. JCESR has assembled a broad multi-institutional team with the expertise and resources to explore the rich and complex beyond lithium-ion battery space and ultimately create next generation battery systems from whole cloth. As JCESR and its colleagues in the small beyond-lithium-ion community plumb the conceptual and material depths of the rich beyond-lithium-ion space, they reduce the scientific and technological risk of failure, attracting other players to the transformative payoff of next generation high performance, low cost batteries.

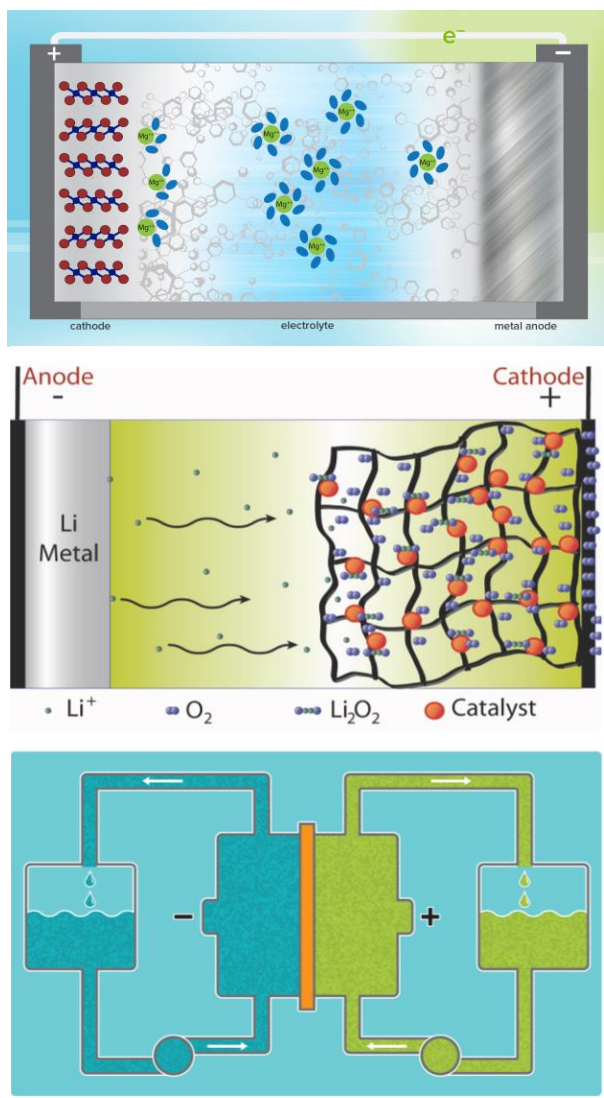

**FIGURE 2.** The three energy storage concepts that JCESR pursues: replacement of singly ionized lithium by doubly charged ions such as magnesium or triply charged ions such as aluminum in multivalent intercalation (upper panel); replacement of intercalation of the working ion at the anode and cathode with higher energy covalent chemical bonds in chemical transformation (middle panel, from Ref 24); and replacement of crystalline electrodes with fluid electrodes in non-aqueous redox flow (lower panel).

## THREE LEGACIES

JCESR intends to produce three outcomes or legacies. The first is a library of fundamental knowledge of the materials and phenomena of electrical energy storage at atomic and molecular levels. This library will be freely available through the literature and open source software, and will inform, inspire and accelerate the work of the broader battery community. Producing and exploiting such a library of fundamental knowledge brings a new dimension to battery research and development. The traditional battery community operates by trial and error, testing a new material such as a better cathode; if it works it is adopted, if it fails it is tossed aside, without asking the question of why it worked or why it failed. JCESR believes that fundamental understanding drives faster and more effective research and development, first in identifying the most promising new materials for batteries, and second in getting the highest performance from those new materials once they are identified.

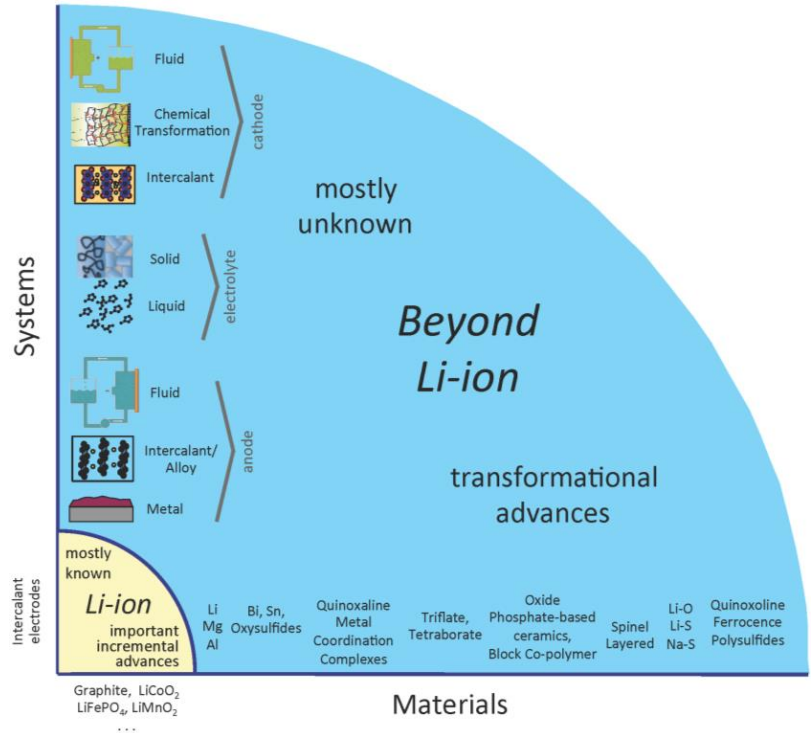

**FIGURE 3.** Beyond-lithium-ion batteries embrace many more storage concepts (vertical axis) and materials for implementation of these concepts (horizontal axis) than present generation commercial lithium-ion batteries. As a commercial technology for more than two decades, present generation lithium ion batteries are technically well understood and still capable of significant but incremental advances. In contrast, beyond-lithium-ion batteries are largely unexplored and promise transformative advances.

JCESR's second legacy will be delivery of two prototypes, one for transportation and one for the grid, that, when scaled to manufacturing, are capable of delivering five times the energy density at one fifth the cost of present generation commercial batteries. These two prototypes for portable and stationary applications will differ in capacity, charging rates and operating environment, but they may be based on the same library of fundamental knowledge produced in the first legacy. JCESR's contract period is five years, a short time to achieve such transformative goals. Nevertheless, JCESR has deliberately chosen not to diminish its vision or replace its mission with less aggressive outcomes that carry lower risk but are not transformative. High performance, lower cost next generation electricity storage is an essential and central lynchpin for next generation energy technology.

JCESR's third legacy is a new paradigm for battery research and development that will accelerate the pace of discovery and innovation and reduce the time from conceptualization to commercialization. The new paradigm, illustrated in Fig. 4, integrates discovery science, battery design, research prototyping and manufacturing collaboration in a single highly interactive organization. In JCESR, these four functions work in close cooperation, unlike the traditional battery community where each function is typically carried out by a separate research organization in a separate location by different experts with differing skills, motivations, and focus. The traditional research cycle of discovery, journal publication, conference presentation, and spontaneous inspiration of new collaborations to exploit the discovery typically takes years to complete. These outcomes can be achieved in weeks or months within JCESR. Further, JCESR deliberately pursues all four of its functions simultaneously, exploiting the dynamic interaction and inspiration that each function draws from the others. Research prototyping, for example, reveals battery design issues and discovery science challenges that are passed rapidly to the appropriate functional team for analysis and solution, enriching the scope and accelerating the progress of all areas.

A traditional Edisonian approach, simply trying one new battery combination after another, is too slow to make adequate progress in the large beyond lithium-ion space. JCESR's new paradigm replaces Edisonian science with innovative new tools that map the broad outlines of the beyond-lithium-ion landscape instead of laboriously exploring its details one battery system at a time. These new tools include the Materials Project, the Electrolyte Genome, the Electrochemical Discovery Laboratory and Techno-economic Modeling, illustrated in Fig. 5 in the context of JCESR's new paradigm.

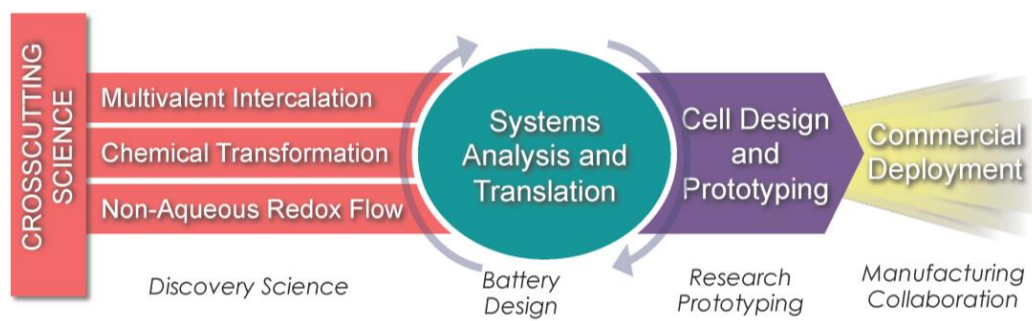

**FIGURE 4.** JCESR's new paradigm for battery research and development, integrating discovery science, battery design, research prototyping and manufacturing collaboration in a single highly interactive organization.

The Materials Project [32] and the Electrolyte Genome [33] use high-throughput computer modeling to discover new working ions, cathodes, anodes and electrolytes, predict their performance and select the most promising candidates before they are made in the laboratory, dramatically reducing discovery time. The Electrochemical Discovery Laboratory uses advanced synthesis and characterization to design and explore high performance electrochemical interfaces, the conceptual and practical heart of electricity storage [34]. Techno-economic Modeling "builds the battery on the computer," projecting performance and cost of proposed battery systems before they are assembled in the laboratory [35].

These innovative tools, combined with rapid exchange of information in JCESR's new paradigm, allow the large unexplored richness of the beyond lithium-ion space to be rapidly surveyed.

# FIRST YEAR ACCOMPLISHMENTS

In its first year, JCESR completed foundational research in all four of its targeted areas: discovery science [36-71], battery design [72-82], research prototyping [83-87] and manufacturing collaboration [88-95]. Most of JCESR's research accomplishments belong to multiple areas, such as the Electrolyte Genome simulation of quinoxaline:$2BF_3$, a discovery science activity prompted by the observation of anomalously high redox activity in prototyping an all-organic redox flow battery (63). This example and other selected accomplishments are highlighted below.

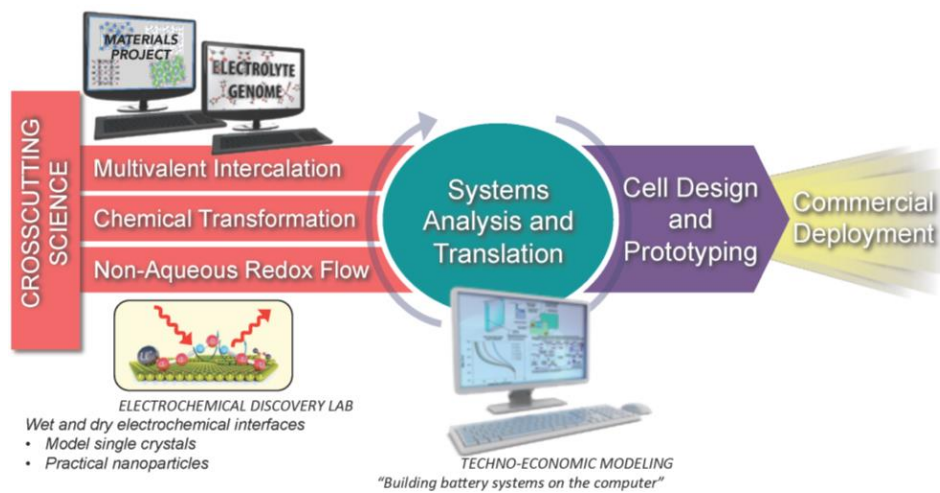

**FIGURE 5.** JCESR's distinguishing tools: high throughput computer simulation of thousands of candidate crystalline electrodes in the Materials Project and liquid organic electrolytes in the Electrolyte Genome, experimental design and characterization of wet and dry electrochemical interfaces in the Electrochemical Discovery Laboratory, and computer simulation of the performance and cost of proposed battery systems in Techno-economic Modeling.

## Discovery Science

*Trace water in lithium-oxygen electrochemistry.* JCESR used the Electrochemical Discovery Lab to discover and quantify a new phenomenon: that trace amounts of water catalyze the conversion of $LiO_2$ superoxide to $Li_2O_2$ peroxide in the lithium oxidation reaction [64,65]. Although water is notorious as a highly active impurity in electrochemical solutions because of its high electric dipole moment, trace water is usually consumed quickly by reaction and hence plays a negligible role in affecting the outcomes of electrochemical processes. In the lithium oxidation cycle shown in Fig. 6, however, water is not consumed, allowing trace amounts to leverage large and sustainable effects. The net effect of trace water in Fig. 6 is the conversion of lithium superoxide to lithium peroxide, leaving the water intact. This surprising result was discovered in systematic experiments of the reaction products in a solution of lithium ions exposed to oxygen in a film of the organic electrolyte DME (dimethoxyethane) on a gold single crystal surface. The water concentration was systematically varied from less than 1 ppm to several tens of thousands ppm. Raman spectroscopy showed that the concentration of lithium peroxide increased strongly as the water concentration rose above 5 ppm, and that the reaction products mimicked those of lithium and oxygen in pure water at concentrations above a few thousand ppm. The first result shows that trace water controls the speed and outcome of the lithium oxidation reaction; the second result shows the extreme activity of water as an electrochemical agent, effectively displacing the organic DME solvent as a player at only a few tenths per cent concentration. Density functional theory simulations confirmed key aspects of the molecular configurations shown in Fig. 6. In battery systems it is impossible to avoid contamination by trace water at several times the 5 ppm level, and usually the trace water concentration is unknown. This research shows that such unavoidable and unquantified trace water contamination can be the determining factor in the electrochemistry of battery charging and discharging. The presence of water, a critical yet uncontrolled component of battery electrochemistry, defies the design of a predictable, reliable and efficient battery cycling strategy.

*Solvation shell of Mg++.* JCESR discovered the solvation shell structure of doubly ionized Mg++ in organic diglyme solvent using a collaboration of Argonne's Advanced Photon Source and the Electrolyte Genome [60,61]. Solvation shells are central to nearly all aspects of battery operation as illustrated in Figs. 1 and 2. They control the mobility of the working ion in the electrolyte, chemical reactions of the working ion with the electrolyte, electrode or impurities in the electrolyte, solubility of the working ion in the electrolyte and they mediate the transfer of the working ion from the electrolyte to an intercalation or chemically reactive electrode. A thorough understanding of the structure and dynamics of the solvation shell of the working ion in the electrolyte is fundamental to battery innovation.

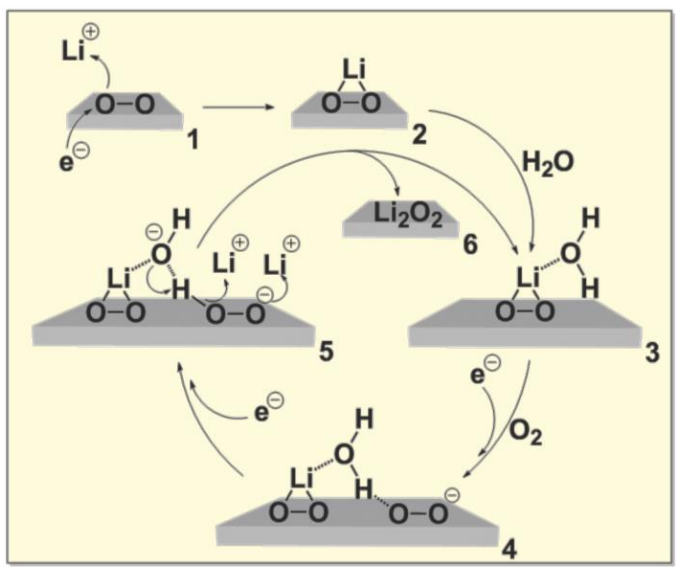

**FIGURE 6**. Trace water controls the kinetics and outcome of the oxidation reactions of lithium. Trace water catalyzes the conversion of lithium superoxide, $LiO_2$ to lithium peroxide, $Li_2O_2$, in the oxidation of lithium. Water is not consumed in the reaction cycle, allowing trace amounts to have large and sustainable effects.

Mg++ solvation shells were created by dissociating Mg(TFSI)$_2$ salt in diglyme solvent. X-ray diffraction of the solution was carried out at the Advanced Photon Source and analyzed with an innovative pair distribution function technique, using multivariate analysis of a series of solutions of many different working ions in the same diglyme solvent to extract the experimental pair distribution function for the Mg++ solvation shell from all other pairs of atoms in the host liquid. In parallel, the Electrolyte Genome calculated the theoretical pair distribution function from molecular dynamics of the dissolved Mg(TFSI)$_2$ salt in diglyme solvent. The theoretical pair distribution function identified the atom pairs producing peaks in the experimental pair distribution function, providing a comprehensive interpretation of its meaning. The positions of the theoretical peaks did not, however, exactly match the experimental peaks, providing a caliper for refining the interaction potentials of the molecular dynamics simulation to produce an exact match of the experiment and theory. Figure 7 shows the comparison of the experimental and theoretical pair distribution functions and the Mg++ solvation shell they represent. A key finding of this new technique is the presence of TFSI- anions in the Mg++ solvation shell, a feature later found in Electrolyte Genome simulations to be common in solvation shells of doubly charged ions in organic solvents. This behavior contrasts sharply with solvation of singly charged lithium ion electrolytes, where the Li+ and anion solvation shells are well separated and distinct. Apparently the strong electrostatic attraction of the Mg++ cation and TFSI- anion overwhelms the weaker interaction of Mg++ with the dipole moment of neutral diglyme molecules.

The presence of anions in the solvation shell of Mg++ reduces its positive charge, significantly affecting mobility and potentially affecting other behavior such as chemical reactions with other constituents of the electrolyte and with electrodes and the transfer of the Mg++ ion from the liquid electrolyte to an intercalated state in a crystalline electrode. This collaboration of x-ray diffraction experiments with Electrolyte Genome simulations opens a new door to systematic characterization and understanding of solvation shell structure in a wealth of working ion-electrolyte combinations for next generation battery technologies.

*Multivalent intercalation.* Implementing the multivalent intercalation energy storage concept illustrated in Fig. 2 requires finding new combinations of anode, electrolyte, cathode and multiply charged working ion that are mutually compatible. Only one such working combination has been demonstrated [96], Mg metal anode and working ion with Grignard electrolyte and Chevrel phase Mo$_6$S$_8$ cathode, illustrating the difficulty of the challenge.

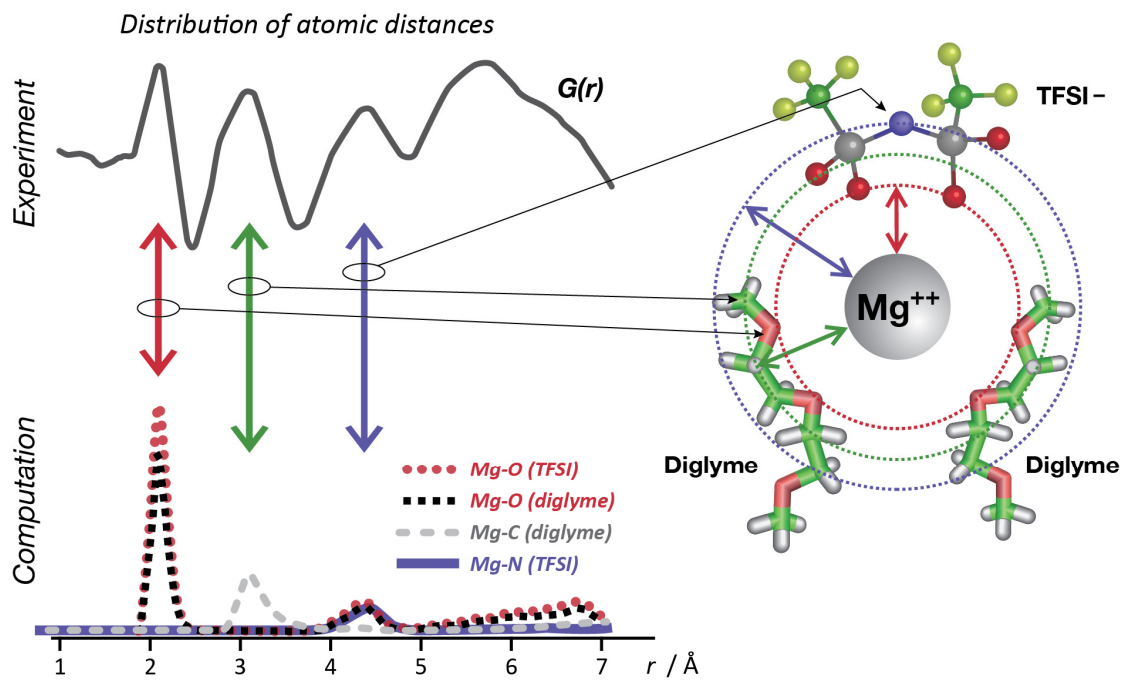

**FIGURE 7.** The solvation shell of Mg++ in diglyme solvent, arising from the dissociation of Mg(TFSI)$_2$ salt, from a collaboration of experimental x-ray diffraction at Argonne's Advanced Photon Source and theoretical simulation in the Electrolyte Genome.

A major part of the challenge is identifying compatible working ion-cathode combinations, which must exhibit high mobility and capacity of the working ion in the intercalation host and stability against disruptive phase transitions as the working ion intercalates or de-intercalates. JCESR addresses the challenge of identifying mutually compatible pairs of multivalent working ion and intercalation host cathode with high throughput simulations [32]. A given working ion-intercalation cathode pair is simulated for several different properties, as illustrated in Fig. 8. The relatively rapid calculation of equilibrium properties is done first using density functional theory, simulating operating voltage, intercalation capacity and phase stability. Non-equilibrium properties such as working ion mobility require more elaborate and slower calculations (a rapid version of mobility simulation is under development and will soon be deployed). The most promising pairs of multivalent working ions and intercalation host cathodes are then selected for experimental synthesis and characterization. This process allows rapid systematic mapping of the vast beyond-lithium ion space, replacing slow experimental synthesis and characterization with fast computer simulation. To date 1800 pairs of multivalent working ions and intercalation host cathodes have been surveyed, with 30 pairs selected as sufficiently promising for further study. Such large scale systematic simulation of the beyond lithium ion space would have been impossible only a few years ago.

The large database of simulated properties allows systematic trends to be isolated. Simulations have shown that the trivalent working ions Al and Y are compatible with a limited number of intercalation hosts that are stable in both the charged and discharged state. Intuitively, it makes sense that it is difficult, but clearly not impossible, to find structures that can accommodate the removal and insertion of the high charge on trivalent cations without yielding to a strong thermodynamic driving force for decomposition. Divalent Mg and Ca are well represented with many good candidates in a voltage range between 2.5 V and 4 V and capacity of 100-200 mAh/g. Zinc is well represented too, but in a significantly lower voltage range (0-3 V). Of the 30 pairs of working ions and cathodes chosen for further study, some are being synthesized experimentally and others are being submitted for more extensive mobility evaluation by simulation.

*Liquid organic electrolytes*. Although liquid organic electrolytes are central to the beyond lithium-ion challenge, they have received far less attention by genomic-style simulation than crystalline materials. JCESR developed a new program, the Electrolyte Genome, to bring broadly based high throughput simulation to the prediction of key electrolyte properties such as the redox operating potential window, stability of liquid electrolyte solvents against detrimental side reactions and solubility of redox-active materials in liquid organic solvents [33]. The Electrolyte Genome seamlessly integrates *ab initio* molecular and classical molecular dynamics simulations to explore single molecules, chemical motifs, and bulk electrolyte structure and performance. This data-driven approach identifies trends and provides insights to guide and accelerate electrolyte design and discovery for next generation batteries.

The Electrolyte Genome tiers screening criteria according to importance for battery technology and calculational speed. For most battery applications, redox operating potential, solubility and stability are of primary and broad interest, while other more specific criteria such as reaction kinetics and pathways with electrode and electrolyte materials and with salt dissociation products require more specialized codes. The tiered screening criteria are illustrated in Fig. 9.

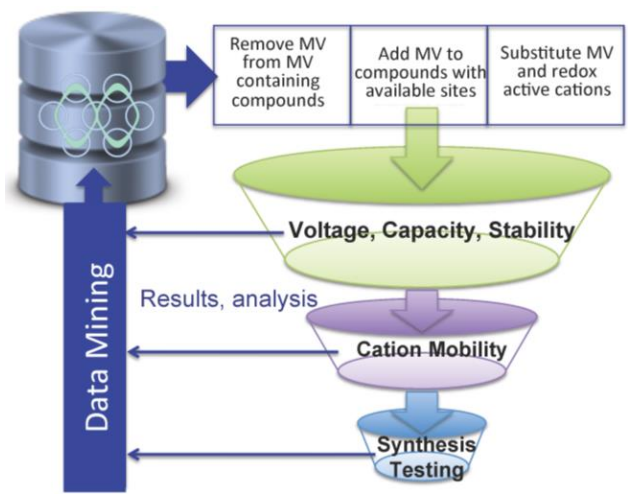

**FIGURE 8.** Computational multivalent working ion-intercalation host cathode screening and learning process.

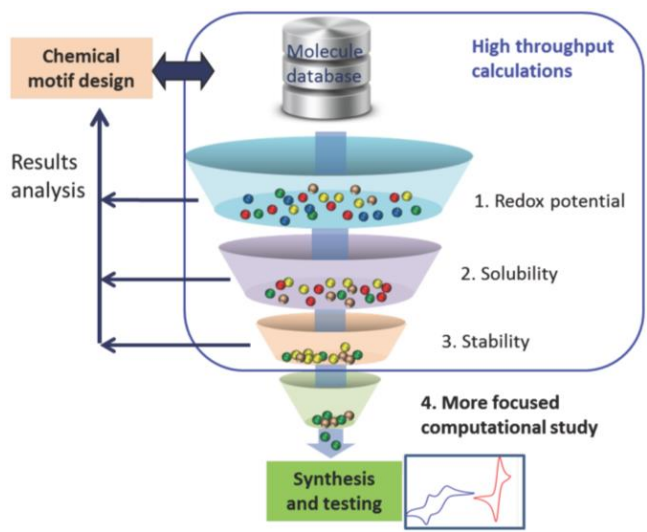

**FIGURE 9**. Tiered screening criteria for the Electrolyte Genome.

As of this writing the Electrolyte Genome has simulated the ionization potential (IP) and electron affinity (EA) of more than 4800 molecules by *ab initio* density functional theory methods and assimilated the results into the Materials Project Database. One intriguing outcome is the discovery that the ionization potential and electron affinity of potential redox-active molecules such as anthraquinone and quinoxaline for use in flow batteries can be tailored by adding functional groups to specific locations on the base molecule [33]. For quinoxaline, the Electrolyte Genome found that the most influential locations for altering the ionization potential of the highest occupied molecular orbital (HOMO) and electron affinity of the lowest unoccupied molecular orbital (LUMO) are distinct and well-separated as illustrated in Fig. 10. Because the two positions are spatially well-separated, the ionization potential and electron affinity can be systematically and independently varied over ranges of at least 2 V by adding specific functional groups at the specified locations, enabling rational design of organic active redox molecules for non-aqueous flow battery applications.

*New electrochemically active Lewis acid-base adducts.* The Electrolyte Genome used density functional simulation to investigate the anomalously high redox activity of quinoxaline derivatives in electrolytes with $LiBF_4$ salts, a feature discovered in experimental prototyping of all-organic flow batteries. Simulations revealed that $BF_3$, a salt decomposition product from $LiBF_4$, reacts spontaneously with quinoxaline to form a Lewis acid-base adduct, quinoxaline:$2BF_3$, with the $BF_3$ units attaching to the nitrogen sites in quinoxaline. The new adduct undergoes two redox events between 2.4 – 3.2 V vs. $Li/Li^+$, about 1 – 1.5 V higher than prior observations of quinoxaline redox activity in non-aqueous media. Direct synthesis and testing of a bis-$BF_3$ quinoxaline complex further validates the assignment of redox active species and demonstrates ~10-fold increase in redox current as compared to unmodified quinoxaline in $LiBF_4$-based electrolyte. The accidental discovery and high redox activity of this adduct suggests the existence of large and promising new classes of Lewis acid-base compounds that might be synthesized readily from heterocycles like quinoxaline with electron-deficient Lewis acids such as $BF_3$.

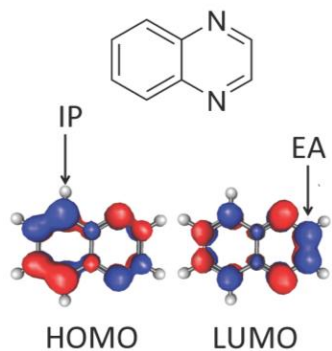

**FIGURE 10.** The double ring heterocycle quinoxaline (upper) and the most influential locations in its structure for altering the ionization potential of the HOMO and electron affinity of the LUMO (lower), as revealed by the Electrolyte Genome.

## Battery Design

*Materials level performance targets.* JCESR used a novel application of techno-economic modeling to infer or "reverse design" materials level voltage and capacity thresholds from system level performance targets. These materials level thresholds establish performance "floors" for discovery science teams looking for new anodes, cathodes and electrolytes for multivalent intercalation, chemical transformation and non-aqueous redox flow storage concepts. These materials performance thresholds provide a selection filter for the output of the Materials Project and Electrolyte Genome simulations of crystalline electrodes and liquid organic electrolytes, identifying those with the potential to meet JCESR's aggressive goals and rejecting those unlikely to be effective. These "reverse designed" materials performance thresholds are quite selective; they typically eliminate a large fraction of the thousands of materials simulated by the Materials Project and Electrolyte Genome, providing a valuable first cut in mapping the beyond lithium-ion landscape. This coupling of system level techno-economic modeling and materials level genomic simulation illustrates the value of strong interaction between battery design and discovery science within JCESR. This kind of interaction significantly accelerates the pace of new materials discovery and would not have been possible only a few years ago.

*System level performance simulations.* Techno-economic modeling provides comparisons of "best case" system level battery performance of battery technology concepts pursued by JCESR, including lithium-oxygen, lithium sulfur, multivalent intercalation and redox flow batteries. [80-82] These "best case' scenarios assume that the most difficult science challenges, such as reversing the lithium-sulfur reaction and achieving reliable cyclability of metal anode stripping and plating, have been overcome. They provide valuable guidance on the research directions with the highest potential impact. Some of these comparisons are shown in Fig. 11, where the system level energy densities of lithium-oxygen batteries are compared to commercial lithium ion batteries for transportation and to proposed lithium-ion batteries with advanced intercalation cathodes and with silicon and lithium metal anodes [80]. The techno-economic model goes well beyond the more common "theoretical energy density" analyses that include the electrochemically active materials only but not the non-active materials such as housing, current collectors, membranes and, for lithium oxygen, the apparatus for purifying atmospheric oxygen or storing and recycling purified oxygen. The techno-economic model includes these essential non-active components in a full system analysis. A remarkable outcome of this study is the significant reduction in the theoretical energy density of lithium oxygen batteries shown in the inset to Fig. 11 compared to system level energy density shown in the main panel. Reductions of as much as a factor of ten can be seen, diminishing much of the appeal suggested by simple active-materials-only estimates.

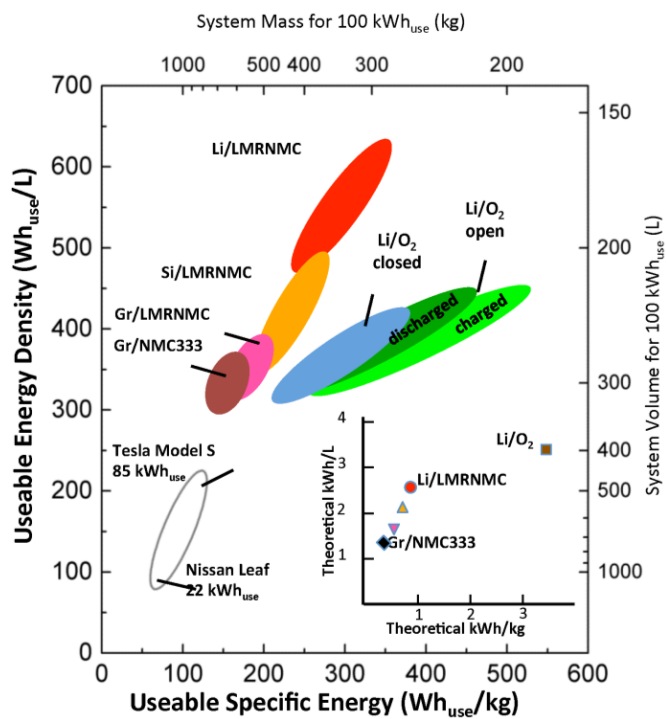

**FIGURE 11.** The useable volumetric energy density and gravimetric specific energy for various batteries, assuming science challenges are overcome, such as reversing the lithium-oxygen discharge reaction and protecting pure metal anodes under repeated stripping/plating.

The major factor diminishing theoretical active-materials-only performance of lithium oxygen batteries is the weight and volume of the apparatus required to remove carbon dioxide and moisture from atmospheric oxygen, or that of the pumps and storage tanks to store and recycle previously purified oxygen. In contrast, the advantage of implementing advanced anodes and cathodes, especially a pure metal lithium anode (red bubble in Fig. 11) is dramatic, potentially exceeding the performance of lithium oxygen on a volumetric basis. It must be remembered that these "best case" scenarios assume that difficult science challenges (such as reversing the lithium-oxygen discharge reaction and protecting a pure metal lithium anode) are overcome, and thus represent potential rather than actual performance. Nevertheless, they provide a new level of insight into best case system level performance that informs strategic research decisions.

## Research Prototyping

*Infinite current collectors.* JCESR developed and submitted patent applications on two complementary flow battery concepts, the Gravity Induced Flow (GIF) cell and the infinite current collector. Conventional current collectors in flow batteries are stationary porous carbon felts or rough metal surfaces that exchange electrons with the liquid electrolyte over a relatively small exposed area and only when the liquid is actively pumped through the carbon felt or past the metal surface. In contrast, the infinite current collector is a diffusion-limited aggregation of nanoscale conductor particles that imparts mixed electronic-ionic conductivity to redox solutions and moves with the electrolyte flow, forming dynamic electrode networks that break apart and self-heal after shear [78,79]. Figure 12 illustrates the basic principles of the infinite current collector.

The infinite current collector extends electrochemical activity throughout the volume of the liquid electrodes compared to the relatively small stationary exposed area of conventional current collectors. In demonstrations with a lithium polysulfide flow cathode, the measured capacity with an infinite current collector was five times that of the same battery with a conventional current collector, owing to more intimate contact and effective electronic exchange with a greater fraction of the liquid. An unexpected benefit is the low pressure required to push the liquid electrolyte through an infinite current collector: 300 times smaller than the pressure needed for a conventional carbon felt current collector with the same flow rate. The infinite current collector concept can be applied to flow batteries of any design and promises to enable higher energy density, greater capacity and lower system cost.

*Gravity Induced Flow cells.* A second innovation introduced by JCESR significantly simplifies the design of conventional flow cells. In the Gravity Induced Flow (GIF) cell, the cathode and anode tanks are mounted on a tilting table straddling the reaction plate or "stack", as illustrated in Fig. 13 [79]. The flow rate in a GIF cell is controlled by the tilt angle, which is adjusted by a small motor, eliminating the more elaborate pumps of a conventional flow cell, lowering cost and reducing maintenance and power requirements. The GIF cell design is fully compatible with the infinite current collector concept described above; the combination of these two innovations has the potential to introduce a new paradigm for design and operation of flow cells. The GIF cell was demonstrated with lithium polysulfide active materials in tetraglyme (TEGDME) solvent operated over several tens of cycles. The demonstration revealed that the flow is non-Newtonian and controlled by many parameters including tilt angle, viscosity, concentration of suspended particles and slipperiness of the liquid-cell interface (which can be tuned with a variety of surface treatments). GIF cells offer the compelling opportunity to design flow parameters for any of several outcomes: high energy density, high capacity, high power or low cost.

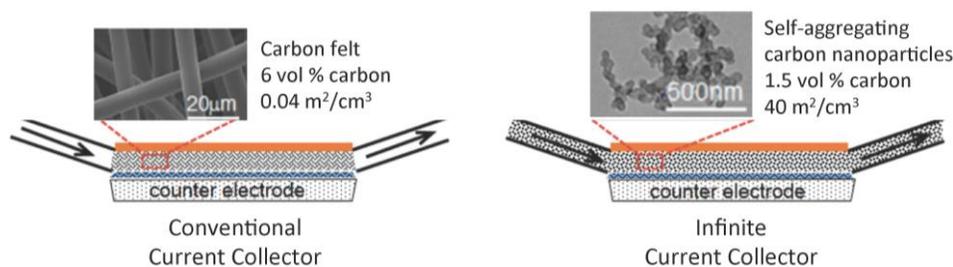

**FIGURE 12**. Conventional and Infinite Current Collectors, adapted from Ref 78 with permission.

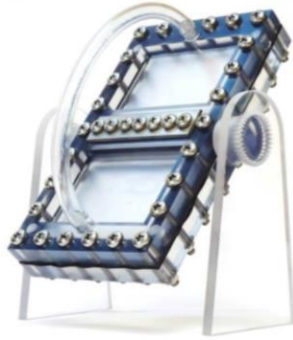

**FIGURE 13**. The Gravity Induced Flow (GIF) cell, where anode and cathode liquid solutions or suspensions in the tanks flow past the reaction plate or "stack" under the influence of gravity. A small motor (not shown) replaces the pumps that drive the liquids in a conventional flow battery, adjusting the tilt of the GIF cell and controling the flow rate.

## Manufacturing Collaboration

*Battery Technology Readiness Levels.* JCESR introduced the concept of Battery Technology Readiness Levels (BTRL) [88] to describe the progression of battery technology development from discovery science through conceptual design, testing, prototyping of research and proof-of-concept cells to commercial product prototyping, illustrated in Fig. 14. The BTRL concept and chart were developed with input from industrial, military and NASA partners and collaborators including Johnson Controls, NASA's Glenn Research Center and the Army's Tank Automotive Research Development and Engineering Center (TARDEC). The BTRL diagram codifies stages of development starting with basic science breakthroughs in understanding materials and phenomena and continues through identification of promising classes of new materials, testing of compatibility and operation in half-cells consisting of anode-electrolyte or cathode-electrolyte combinations, identification and testing of compatible full cell systems comprising complete anode-electrolyte-cathode combinations, design of cells capable of meeting specific performance metrics, and translation to a industrial firm or consortium that will launch prototyping of larger scale systems as models for manufacturing and commercialization.

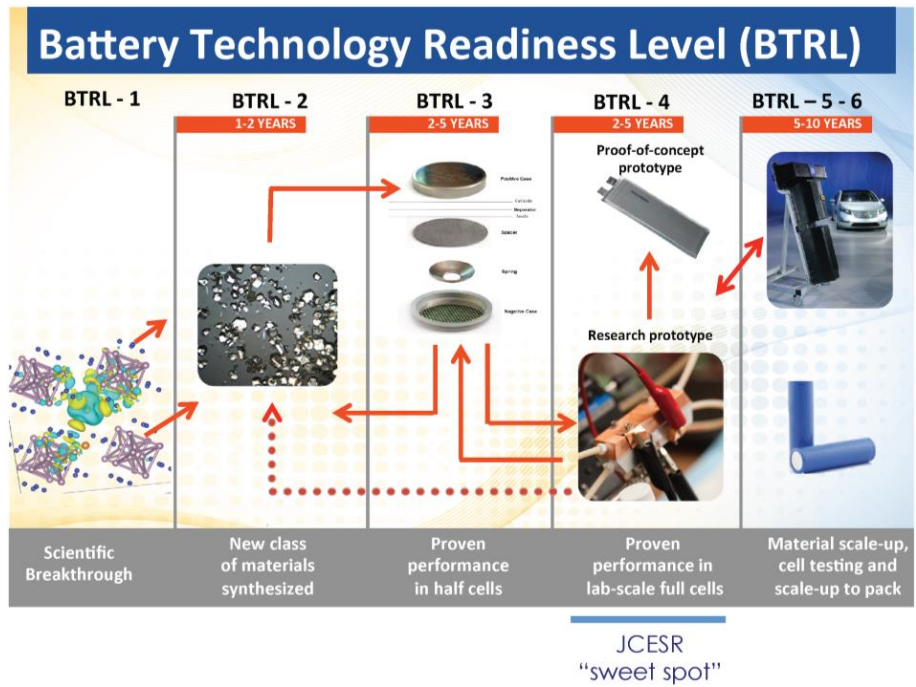

**FIGURE 14.** Battery Technology Readiness Levels introduced by JCESR to codify the stages of battery technology development from discovery through translation to commercial prototyping.

JCESR's target for prototyping is captured as BTRL-4, research or proof of concept prototypes that demonstrate the potential to achieve factors of five improvement in performance and reduction in cost that will drive transformative change in transportation and the electricity grid. Laboratory characterization of JCESR's prototypes provides materials and system level performance data that are inputs to techno-economic models of the cost and performance of manufactured batteries scaled up from JCESR's research and proof-of-principle prototypes. This projection to manufactured performance using techno-economic modeling is a critical step in JCESR's prototyping agenda, supporting or opposing a decision to move from the laboratory to commercial-scale prototyping and manufacturing.

*JCESR Affiliates.* JCESR reaches out to the broader battery community through its Affiliates program for private corporations, startup companies, venture capitalists, trade associations, research universities, Energy Frontier Research Centers and other organizations with a research and development interest in JCESR's program, progress and outlook. The Affiliates include potential science and technology collaborators, strategic planners for the trajectory, ultimate value, and eventual impact of beyond lithium ion batteries in transportation and the grid, policy analysts for the implications of next generation storage for the broader energy system and society, and large and small companies who may wish to license JCESR's intellectual property. [89,90] JCESR convened its Affiliates for a day-long meeting in spring 2014, attracting 85 participants representing 50 organizations. [91,92] At its launch in January 2013 JCESR counted 45 Affiliates; that number has grown to 75 at the time of this writing. As part of its affiliates program, JCESR organizes regional events to identify and analyze key challenges, opportunities and outcomes of commercialization of next generation high performance, low cost electricity storage. [93,95]

*Venture Capital Advisory Council.* JCESR convened its Venture Capital Advisory Council consisting of thought leaders and strategic planners in the energy community.[94] This group has met three times in the San Francisco Bay area, to consider questions such as market opportunities and commercialization strategy for next generation electricity storage. The strategic thinking and planning of this group for the eventual rollout of next generation energy storage is central to JCESR's new paradigm for battery research and development.

## PERSPECTIVE

In its first year, the JCESR partnership has moved rapidly from launch to full operation and is producing groundbreaking research in discovery science, battery design, research prototyping and manufacturing collaboration. The ultimate battery may combine several of the storage concepts that JCESR is exploring. It might, for example, use multivalent ions intercalated into the anode and strong chemical bonds at the cathode, or a crystalline anode and a liquid cathode. JCESR has begun to map the opportunities in the vast, rich and largely unexplored beyond-lithium-ion space, using its distinguishing tools of high throughput modeling of crystalline materials through the Materials Project and liquid organic electrolytes through the Electrolyte Genome, design and evaluation of electrochemical interfaces through the Electrochemical Discovery Laboratory, and modeling and evaluation of system-level battery performance and cost through Techno-economic modeling. JCESR maps the rich beyond-lithium-ion space, identifying and pursuing the most promising beyond-lithium-ion electricity storage opportunities as it builds its three legacies: a fundamental understanding of the materials and phenomena of energy storage at atomic and molecular levels, research and proof-of-principle prototype batteries for transportation and the grid capable of meeting JCESR's aggressive performance and cost goals, and a new paradigm for battery R&D that integrates discovery science, battery design, research prototyping and manufacturing collaboration in a single, highly interactive organization.

## ACKNOWLEDGMENT

This work was supported as part of the Joint Center for Energy Storage Research, an Energy Innovation Hub funded by the U.S. Department of Energy, Office of Science, Basic Energy Sciences. The submitted manuscript has been created by UChicago Argonne, LLC, Operator of Argonne National Laboratory ("Argonne"). Argonne, a U.S. Department of Energy Office of Science laboratory, is operated under Contract no. DE-AC02-06CH11357.

Except where otherwise noted, all figures originate within the Joint Center for Energy Storage Research and are in the public domain.